\documentclass[11pt]{article}
\usepackage[utf8]{inputenc}
\usepackage[T1]{fontenc}
\usepackage{lmodern}
\usepackage{authblk}
\usepackage{mdframed}
\usepackage{amsmath, amssymb, amsthm}
\usepackage{graphicx}
\usepackage{booktabs}
\usepackage{float}
\usepackage[margin=1in]{geometry}
\usepackage{placeins}
\usepackage{enumitem}
\usepackage{fancyhdr}
\usepackage{caption}
\usepackage{hyperref}
\hypersetup{hidelinks}

\title{\textbf{LEARNING BY GAMING, CODING AND MAKING WITH EDUMING: A new approach to utilising atypical digital games for learning}}
\author[1]{Dr. Stefan Pietrusky}
\affil[1]{Heidelberg Center for Digital Humanities (HCDH), Heidelberg University}
\affil[1]{\texttt{stefan.pietrusky@uni-heidelberg.de}}
\affil[1]{Down Church Studios}
\date{\today}

\makeatletter
\let\blx@rerun@biber\relax
\makeatother

\begin{document}

\maketitle

\begin{mdframed}[linewidth=0.5pt, linecolor=black, frametitle={Abstract}, frametitlealignment=\centering]
\noindent
Papert's constructionism makes it clear that learning is particularly effective when learners create tangible artifacts and share and discuss them in social contexts. Technological progress in recent decades has created numerous opportunities for learners to not only passively consume media, but to actively shape it through construction. This article uses the EDUMING concept to present a new method to simplify the development of digital learning games and thus support their integration into learning situations. A key difference between the concept and established ideas such as game-based learning, gamification, serious games, etc. is that games are not closed and are consumed passively, but can also be actively developed by users individually by modifying the source code with the help of an IDE. As part of an empirical study, the usability of the game “Professor Chip's Learning Quest” (PCLQ) is recorded, as well as previous experience with digital learning games and the acceptance and motivation to use new technologies. The purpose of this article is to test the PCLQ digital learning game, developed according to the EDUMING concept, as part of an exploratory study regarding its usability, acceptance and suitability for use in schools. The study is intended as a first empirical approach to practical testing of the concept.
\end{mdframed}

\section{THE EDUMING CONCEPT}
There are various concepts, each with its own approach, for how digital technologies and games can be used for learning purposes. Gamification involves applying typical game elements (badges, rankings, points, loot boxes, etc.) in a non-gaming context. The goal is to promote participation and motivation \cite{1}. This is not a complete game, but rather simply implements mechanics familiar from games into learning platforms or learning apps.

The goal is to increase motivation and productivity through reward systems and competitive aspects. When a game has been developed specifically for educational purposes or an existing one is used in this context, it is referred to as Game-Based Learning (GBL) \cite{2}. In contrast to gamification, this concept not only increases motivation, but also directly imparts knowledge and skills through the game. To emphasize that digital games are meant, it is sometimes also referred to as Digital Game-Based Learning (DGBL). In addition to gamification and DGBL, there are also serious games. Like GBL, these are complete games that were developed primarily for a serious purpose and whose focus is not on entertaining the user, but on solving problems or training and simulating scenarios \cite{3}. Serious games are frequently used in medicine (e.g., surgical training), the military (e.g., strategic planning), and also in environmental education. In this context, edutainment and simulations are also worth mentioning. Edutainment, a combination of education and entertainment, aims to present educational content in an entertaining way \cite{4}. However, compared to serious games and GBL, edutainment pursues a less structured learning approach, meaning it is not based on a methodological-didactic structure to work towards specific learning objectives. A simulation is generally understood to be a digital replication of real systems or processes for analyzing complex scenarios \cite{5}. Simulations are therefore a component of serious games. EDUMING differs from the aforementioned concepts in many respects. The EDUMING concept is centrally based on Seymour Papert's constructionism, according to which learners experience in-depth learning through the active construction of concrete and sharable artifacts \cite{6} \cite{7}. In contrast to well-known constructionist approaches such as Kafai and Burke (2015) \cite{8}, which primarily focus on the complete development of new games, EDUMING deliberately builds on existing game templates to make it significantly easier to get started with creative adaptations. A digital learning game based on the EDUMING concept is not a complete game in the sense of a closed structure with a predefined ending. An EDUMING game is available as an adaptable template that can be further developed using an integrated development environment (IDE). This means that a game based on this concept can be used in various contexts. In an EDUMING game, the focus is neither on learning nor on playing. The focus is defined by the individual user. Every EDUMING game template contains a standardized learning mechanic that can be adapted or completely ignored. The development time and costs are lower compared to traditional applications because the games are based on templates already provided by various IDEs such as GameMaker Studio 2, Unity, or Unreal Engine. Of course, templates for EDUMING games can also be developed directly, depending on the financial support available. To make new didactic concepts realistically usable, considering the financial and time commitment is important. Using tools like GMS2, EDUMING explicitly addresses these factors, ensuring practical pedagogical use.

The goal of an EDUMING game is either to entertain or to impart knowledge through the game itself. However, knowledge is also expanded by giving the user the opportunity to further develop the actual gaming experience. The user therefore learns not only by using the game, but also by further developing the game in the spirit of "learning by making." Another criterion is that AI can be used in the further development of the game to increase development time, costs, and replay value by automatically generating tasks with the help of an LLM and implementing them into existing learning mechanics, or by adapting or optimizing the code (e.g., FPS, frame rate). EDUMING games can also use game types that are not typical for learning (e.g., shooters), and they must also be playable on mobile devices. This distinguishes the EDUMING concept significantly from the already familiar terminology, which completely ignores the further development of the actual application and leaves users passive at this level. In keeping with Papert's constructionism, the EDUMING concept therefore provides a particularly effective learning opportunity. Due to the basic structure of the EDUMING game, the learner can build on it to create a tangible artifact to share and discuss in social contexts \cite{7}.

Depending on how the user wants to adapt or further develop the game, more or less extensive programming knowledge is required. For general adaptation of the game (e.g., visualizations, sounds, etc.), instructions are made available with the template via GitHub repositories. For more complex changes, for example, if the user wants to incorporate new mechanics, the basic knowledge necessary for adapting the game is also conveyed with the help of instructions. Through this approach, the EDUMING concept, in keeping with Papert's philosophy, promotes creative problem solving \cite{6}. A game that corresponds to the EDUMING concept is presented below.

\section{THE IDE GAMEMAKER STUDIO 2}
\begin{figure}[htbp]
    \centering
    \includegraphics[width=1.0\textwidth]{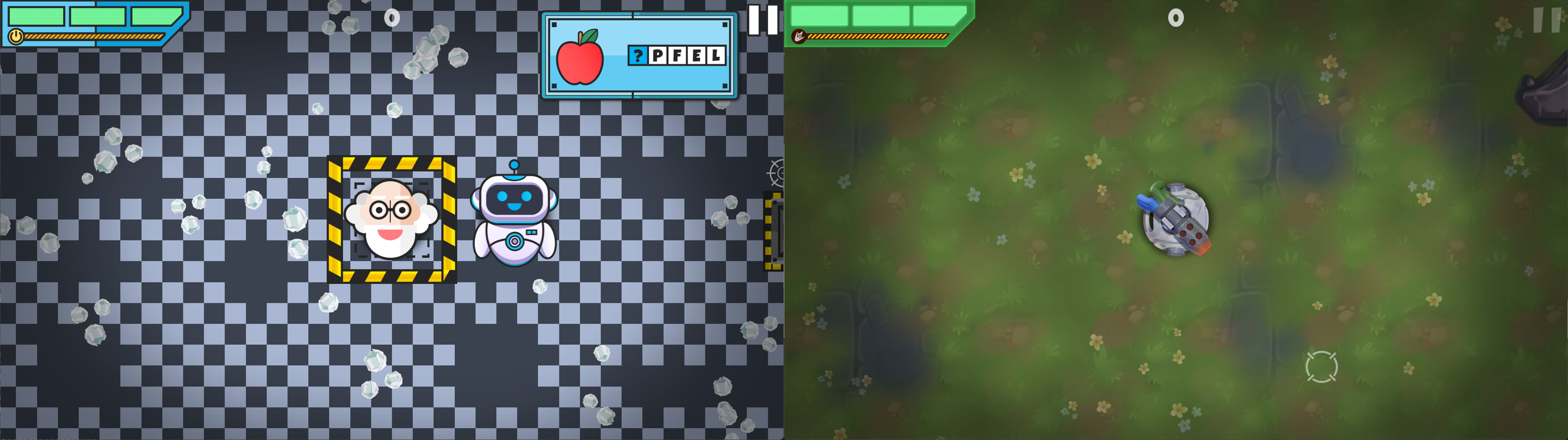}
    \caption{Professor Chips Learning Quest EDUMING Game (left); Twin Stick Shooter Template YoYo Games (right)}
    \label{F1}
\end{figure}

Recent studies demonstrate that both gamified learning environments \cite{9} \cite{10} and adaptively designed learning games \cite{11} can have significant positive effects on learning outcomes and motivation. Similarly, Vlachopoulos and Makri (2017) \cite{12}  confirm the effectiveness of digital games and simulations in higher education in a comprehensive analysis. The EDUMING concept integrates these findings by combining learning-enhancing gamification elements with constructive, adaptive, and creative game mechanics.
To make the EDUMING concept more understandable, it is explained practically using a 2D game. The game is called "Professor Chip's Learning Quest" (PCLQ) and is a so-called arena shooter developed by the indie developer studio Down Church Studios. The IDE used is GameMaker Studio 2 (GMS2) from the developer YoYo Games. The template is also provided via this application. The template is called TwinStick Shooter and is a classic arena shooter in which the level or arena is generated based on a percentage (see Fig.~\ref{F1}). The player starts in the center of the arena, into which enemies enter at specific points. The player must destroy these enemies by shooting at them and thereby earns points. The enemies appear in waves. The higher the wave, the more enemies appear. In the template there is only one enemy type and the difficulty level cannot be adjusted. The player has a total of three lives, meaning they can be hit three times. The projectiles the player fires can be reloaded. When the player is hit three times, the game is over, or when wave 10 is completed. The points achieved are saved and can be viewed using a simple high score function in the main menu.

\pagebreak

In the template, you'll find documentation in the Notes section of the Asset Browser, which briefly explains the various components and their functions. This section also includes initial suggestions for how the game could be further developed. Concrete examples include adapting the level design with a custom tile set, increasing the difficulty by allowing enemies to take more damage, or reducing the player's fire rate. Implementing power-ups and integrating local multiplayer are also possibilities. The game has been adapted in various areas for the EDUMING concept described above. 

To adapt a game, you need to understand how the IDE it was developed with works. The advantage of GMS2 is that getting started is easier than with other development environments. The reason for this is that all development work takes place in the GMS2 IDE. The code editor is part of the development tool and is therefore directly linked to visual development. With Unity, however, the visual design takes place in the Unity editor, and programming in an external editor (e.g., Visual Studio Code). The GMS2 IDE uses GameMaker Language (GML) as its programming language and is limited to 2D applications. The language is easier to learn than other languages such as C\# or C++, but is restricted to the GM environment. The 2D limitation does not immediately overwhelm beginners, compared to the countless functions offered by applications such as Unity or the Unreal Engine. In addition to programming with GML Code, there is also GML Visual. Here, projects can be adapted using blocks without programming. GMS2 offers various options for publishing games. For the EDUMING concept, the simplest option has been chosen, specifically publishing on GX Games. The reason for this is that, according to constructionism, the learner can make their adapted game (=artifact) available to other learners as easily as possible. The publishing procedure is part of the instructions. GMS2 is therefore a very good introduction to the EDUMING concept.

As mentioned above, GMS2 features integrated resource management, allowing sprites, sounds, and scripts to be organized directly in the editor. Game logic is determined by an event-based system, making development more linear. Sprites define the game's visual representation. Objects determine the game's logic. A sprite is assigned to an object, which is then rendered accordingly. Events are created in an object. Variables with values are defined in a "Create" event. When the object is retrieved, these variables and their values are available as functions. Here's an example to help you understand it better. In the Asset Browser of the TwinStick template, we have the "obj\_player" object. The "spr\_player\_body" object has been assigned as a sprite. When the player appears in the game, their appearance is thus tied to the sprite. Some developers abbreviate objects with a lowercase "o" at the beginning and sprites with a lowercase "s." Others write "obj" (object) and "spr" (sprite). You can choose your own names for the assets, but they should be consistent for clarity. Various events exist in "obj\_player." The "Create" event contains numerous variables that are explained using comments (//). The variable "player\_health" defines the player's health points (HP) as 3. In the "obj\_projectile" object, the "Collision" event in "obj\_player" shows the conditions under which the player's HP is reduced. If a projectile hits a player, their HP is reduced by 1 (other.player\_health--;). This example is intended to illustrate how the programming logic in GMS2 works. Variable values can be adjusted for simple customizations. For more complex extensions, users will need to delve deeper into the code structure. The following chapter will discuss the customizations for the EDUMING game PCLQ and expand on the concept.

\section{EDUMING 1 [PCLQ]}
An important criterion for an EDUMING game is the learning mechanics included by default. To better understand the mechanics, a brief overview of the game's story. Professor Chips was actually setting up his new assistant when his cleaning robots were infected by a computer virus and sucked up all the language chips. Now the chips must be retrieved and delivered as pure data, but only if the correct task is displayed. The learning mechanics for PCLQ work as follows. If the player "obj\_player" hits and destroys an opponent "obj\_enemy" with a projectile, the opponent loses an object "oLetter" whose sprite is randomly selected from an array "var\_missingletter\_sprite = [];". The opponent can therefore lose a language chip of class A, B, C, D, or E. However, the chips are unstable and self-destruct if they are not picked up quickly. If the player picks up a language chip in time, the pure data is displayed above their head as a sprite. Which data, or rather which letter, is currently required is represented by the "oMissingLetter" object. As with "oLetter," the sprite is randomly defined using an array.

\begin{figure}[htbp]
    \centering
    \includegraphics[width=1.0\textwidth]{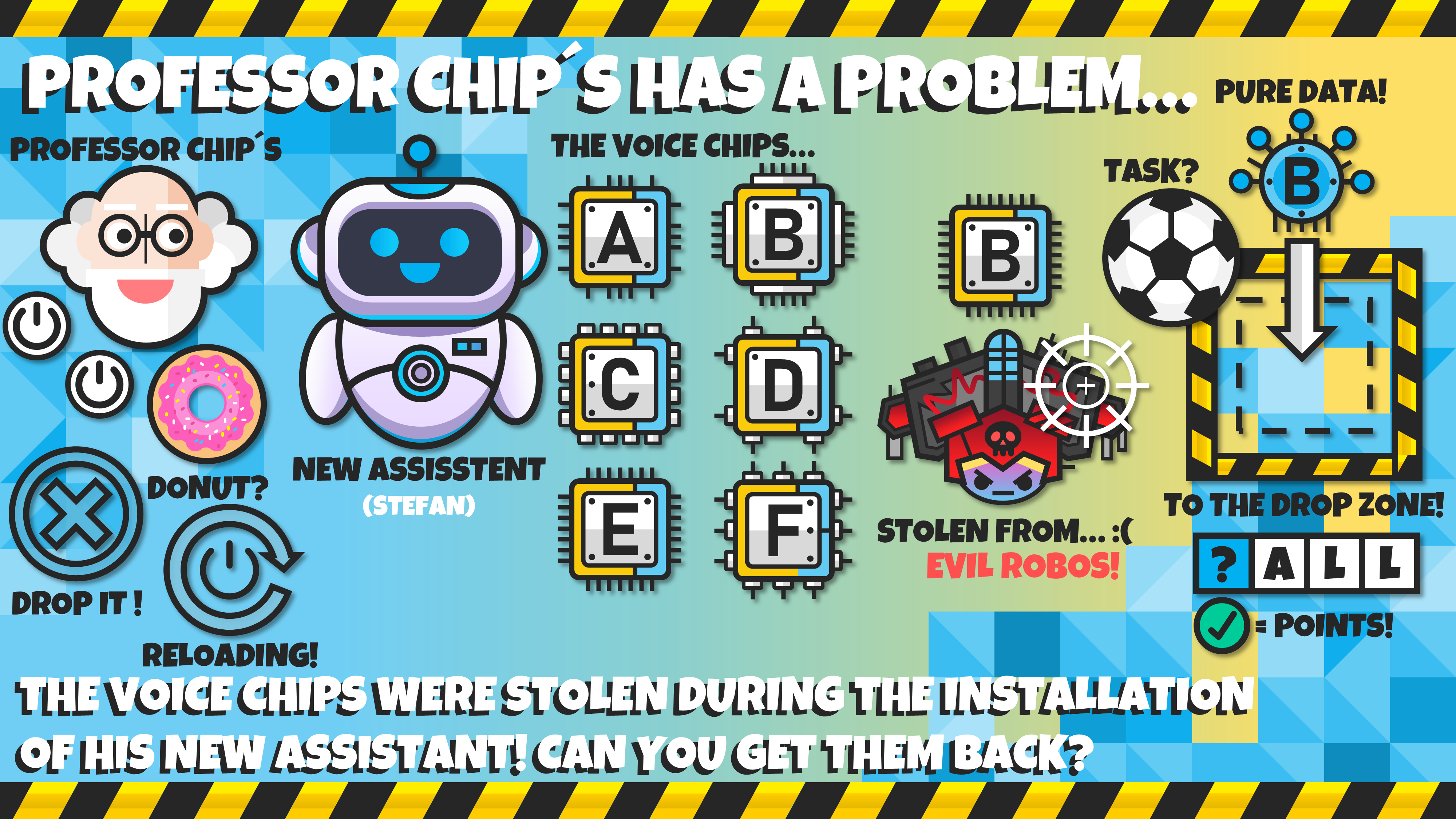}
    \caption{How to play PCLQ EDUMING Game}
    \label{F2}
\end{figure}

So, if the letter A is needed, the player can deliver the data for the letter A to the "oDZ" drop zone, located in the center of the arena, to advance the wizard's configuration. If the correct letter is delivered, the player receives more points than if they completed the game the normal way, i.e., reached the last wave. If a letter is delivered that is not currently required, points are deducted. Regardless of this mechanic, the player still receives points for destroying cleaning robots. If a task is answered correctly or incorrectly, a new one is displayed. The gameplay is explained in the following illustration, which can be viewed in the game via the main menu (see Fig.~\ref{F2}).

\pagebreak

As defined in the EDUMING concept, a game mechanic was built into a template that can easily be expanded to include additional tasks. The various variables and parameters that make up this mechanic can be adjusted using the corresponding objects. PCLQ is about promoting spelling skills. However, the game could also be adapted so that numbers must be submitted instead of letters, or it could be used for a completely different subject area. In addition to revising the entire visual presentation of the original game, a sequence was integrated that can be started via the menu to explain the game principle (seq\_game\_info). Transition effects between the menu and the game were integrated, as well as a script for a screenshake effect that is triggered when the player is hit by a projectile. While the controls on laptops or PCs are controlled using the classic WASD key assignment in combination with a mouse, on smartphones the controls work automatically using the displayed navigation elements in the form of joysticks. In order to empirically validate the EDUMING concept, the following chapter examines the usability of the game PCLQ as well as the acceptance, use and previous experience of the participants in dealing with digital learning games.

\section{MATERIALS AND METHOD}
This study is exploratory in nature and serves to test a first prototype game developed based on the EDUMING concept. The goal is to obtain initial evidence of usability, acceptance, and perceived potential for use in schools.

The ISO 9241-11 standard was used to assess the usability of the digital learning game PCLQ. This standard is used worldwide in usability research and is considered the standard for evaluating interactive systems by measuring effectiveness (goal achievement), efficiency (effort to achieve the goal), and satisfaction (positive user experience) \cite{13}. Acceptance and use are measured using the UTAUT model. UTAUT was selected because it unifies existing models of technology acceptance and is appropriate for this context \cite{14}. In addition to this data, general demographic information such as gender, year of birth, subject combinations, etc., as well as the participants' previous experience with digital learning games, were also collected. The participants were specifically selected from teacher training students who voluntarily completed further training in special education and language didactics in the winter semester of 2024/25. Since the PCLQ game aims to promote spelling skills, the professional expertise of this group is particularly relevant for a pedagogical-didactic assessment. The students were recruited through the voluntary additional qualification program of the Center for Migration Research and Transcultural Pedagogy at the Heidelberg University of Teacher Education, as this program includes teacher training students who are already sensitized to topics such as individualized support and the use of digital learning media in the classroom. The first data collection took place from January 17 to 22, 2025, using an online questionnaire. The second survey was conducted on January 24, 2025, at the Heidelberg University of Teacher Education after the direct testing of the digital learning game. Data were collected using an anonymized questionnaire. The majority of questions, except some demographic and open-ended items, were answered using a five-point Likert scale. 

To accommodate all participants' operating systems, the PCLQ app was available for download as an apk file (Android) and could be used on iOS devices via publication on GX Games. To use the executable file, participants had to put their Android smartphones into developer mode. A brief introduction was given at the beginning of the trial. Permission to conduct the study was obtained from the person in charge of the seminar and the institute management. The students were informed of their rights regarding data processing. Consent was obtained through participation. The following section discusses the results of the first and then the second data collection.

\section{RESULTS}
The first questionnaire was used to record demographic data, previous experience with digital learning games and technology acceptance using the UTAUT model. The second questionnaire recorded user-friendliness with the help of effectiveness, efficiency and satisfaction with regard to the PCLQ game.

\section{RESULTS OF QUESTIONNAIRE 1}
The first questionnaire was completed by a total of 19 participants. Of these, 16 were female and 3 were male. Their birth years ranged from 1993 to 2003, which is why the arithmetic mean is approximately 1999, meaning the participants were on average between 25 and 26 years old. The number of semesters they had already studied ranged from 3 to 13 semesters. On average, participants studied for approximately 8 semesters. One student reported 17 semesters, 10 of which were for teacher training. The participants' subject combinations varied widely (biology, German, English, religion, general studies, sports, technology, and math). Many also indicated a focus on special education, which is understandable, as the participants were recruited from the area of a voluntary additional qualification at the Center for Migration Research and Transcultural Pedagogy. Regarding their experience with digital learning games in the context of their studies, the majority (90\%; n=17) stated that they had little to no experience (M=1.89; s=0.57). In contrast, experience is more widely distributed (M=2.68; s=0.95) and somewhat higher in the private sphere. Some (26\%; n=5) of the participants stated that they had already gained considerable experience. Irrespective of this, half of the participants have had little to no experience so far. Regarding satisfaction with their previous experiences with digital learning games, the majority were satisfied (53\%; n=10), about a third were neutral (32\%; n=6), and only a few participants were dissatisfied or very dissatisfied (15\%; n=3). Satisfaction with their previous experiences thus lies between neutral and satisfied (M=3.32; s=0.89). The following are the results of the questions dealing with the acceptance and use of digital learning games. The majority of participants (79\%; n=15) rated the benefits of digital learning games positively compared to analog learning methods (M=3.84; s=0.69). The majority of participants (84\%; n=16) also agreed with the question that digital learning games effectively improve learners' learning (M=4.11; s=0.57). The idea that digital learning games can support the achievement of teaching objectives was answered with either the highest or second-highest level of agreement on the Likert scale (M=4.37; s=0.50).

The frequently cited advantages of using digital learning games in the classroom can be summarized as follows: They promote learner activation and motivation, offer an interactive and varied learning process, are highly relevant to real-life situations, and have considerable potential to support differentiation and individualized support. Opinions differ regarding the aspects of implementation. Some participants see the use as involving a great deal of effort (21\%; n=4), while others see it as involving none at all (37\%; n=7). Most participants (42\%; n=8) answered the question neutrally (M=2.95; s=1.18). For some participants (19\%; n=3), the opinions of other teachers on the use of digital learning games are an aspect that plays a role in implementation (M=2.63; s=1.01). However, for the majority, the opinions of other teachers play no role (53\%; n=10). Support from school administration, on the other hand, is relevant for the majority (58\%, n=11) of participants (M=3.16; s=0.83). Factors that influence the use of digital learning games include the school's technical equipment (e.g., Wi-Fi or availability of devices such as tablets, etc.), the complexity and quality of the digital offering, and also the learners' prerequisites (e.g., class structure, media literacy, and discipline). The time required for preparation was also mentioned several times. Despite the factors mentioned, there was hardly a clear negative response among participants regarding their willingness to use digital learning games. Only two participants (10\%) answered "neutral." Most participants (90\%) would nevertheless use these media in class (M=4.00; s=0.58). When asked about obstacles or concerns regarding the use of digital learning games in class, the answers were largely similar to those given when asked about the factors influencing their use. Added to this were the lack of and uneven equipment in schools, as well as the potential for distraction from other apps that could be used instead of the actual learning application.

Overall, the data from the first questionnaire show that participants had little experience with digital learning games during their studies. A broader spectrum of opinions emerges in their private lives. Satisfaction with previous experiences is perceived to be predominantly positive. The general attitude toward digital learning games is predominantly optimistic. The obstacles lie primarily in the technical equipment, the potential for distraction, and the additional time and organizational effort required. Nevertheless, the willingness to use digital learning games in the future is largely high. Participants have a positive attitude toward digital learning games and see their benefits, provided the framework conditions (technology, time, and support from the school) are right.

\section{RESULTS OF QUESTIONNAIRE 2}
The second questionnaire was answered by 17 participants. The ISO 9241-11 standard was used to record the user-friendliness of the digital learning game. The effectiveness, efficiency and satisfaction were determined after the application was tested using a five-point Likert scale (strongly disagree [5], disagree [4], neutral [3], agree [2], strongly agree [1]) to enable a more precise and differentiated analysis. The scale was used as specified to calculate the mean values and standard deviation. For the evaluation of the three categories, the weighting was reversed and the following point distribution was used (strongly disagree [1 point], disagree [2 points], neutral [3 points], agree [4 points], strongly agree [5 points]). Effectiveness and efficiency were measured directly by the participants and not by external observation. The following table shows the results of the individual items in the area of effectiveness.

\captionsetup{justification=centering}

\begin{table}[htbp]
\centering
\begin{tabular}{p{9cm} c c}
\toprule
\textbf{Item} & \textbf{Mean} & \textbf{Standard Deviation} \\ 
\midrule
I had no problems navigating through the game. & 2.71 & 1.45 \\
I had no problems starting the game. & 3.00 & 1.17 \\
The visualizations used in the menu are attention-grabbing. & 4.00 & 0.71 \\
The music used in the menu is attention-grabbing. & 4.06 & 0.75 \\
The size of the game window is sufficient. & 4.06 & 1.14 \\
I had no problems moving the character. & 3.59 & 1.28 \\
I knew what I had to do in the game. & 2.41 & 1.00 \\
I had no problems collecting the letters. & 2.88 & 0.93 \\
I had no problems delivering the correct letters. & 2.53 & 1.18 \\
I knew when a letter delivery was correct or incorrect. & 3.53 & 1.28 \\
I understood the overall concept of the game. & 3.24 & 1.03 \\
\bottomrule
\end{tabular}
\caption{Results of the items in the category Effectiveness.\\
Mean values and standard deviations of participant ratings.}
\label{t1}
\end{table}

The table shows that each of the 17 participants had to answer 11 items. According to the point distribution, a maximum of 85 points was possible per item. The maximum number of points for all questions is therefore 935, which means an effectiveness of 100\%. The advantage of this calculation method is that it is not only the success rate that determines the performance in a category. The effectiveness of the digital learning game can be calculated as follows:

\[
\text{Effectiveness (\%)} = \frac{\text{Points per item}}{\text{Total possible points}} \times 100\%
\]

A total of 610 out of 935 maximum possible points were achieved. Based on the equation, it can be said that the effectiveness of the PCLQ digital learning game is around 65\%. Next, the efficiency, i.e. the smoothness with which a task is completed, is determined. The calculation is carried out using the same method as for effectiveness.  The following table shows the list of items in the efficiency section.

\captionsetup{justification=centering}

\begin{table}[htbp]
\centering
\begin{tabular}{p{9cm} c c}
\toprule
\textbf{Item} & \textbf{Mean} & \textbf{Standard Deviation} \\
\midrule
I was able to collect the required speech chips in time before they self-destructed. & 3.06 & 1.39 \\
The display of the required letter helped me react quickly. & 3.35 & 1.17 \\
It was easy to recognize which speech chip was currently required. & 4.29 & 0.77 \\
I immediately understood how the game mechanics worked. & 2.24 & 0.97 \\
I was able to fire the projectiles at the enemies without problems. & 2.94 & 1.25 \\
It was easy to complete the tasks with minimal effort. & 2.71 & 1.05 \\
I did not feel overwhelmed during the game. & 2.65 & 1.27 \\
I rarely had to think long to make the right decision. & 3.18 & 1.13 \\
\bottomrule
\end{tabular}
\caption{Results of the items in the category Efficiency.\\
Mean values and standard deviations of participant ratings.}
\label{t2}
\end{table}

The table shows that each of the 17 participants had to answer 8 items. Again, a maximum of 85 points was possible per item. This results in a total efficiency score of 680 points for all questions. The efficiency of PCLQ is calculated in the same way as the effectiveness. 415 out of 680 points were achieved, which is why the efficiency is 61\%. Satisfaction is calculated last. The following table shows the list of items in the area of satisfaction.

\captionsetup{justification=centering}

\begin{table}[htbp]
\centering
\begin{tabular}{p{9cm} c c}
\toprule
\textbf{Item} & \textbf{Mean} & \textbf{Standard Deviation} \\
\midrule
The game was fun. & 3.35 & 1.32 \\
The game was easy to play. & 2.82 & 1.13 \\
I liked the character in the game. & 3.00 & 1.17 \\
The tasks in the game were easy. & 3.59 & 0.87 \\
The game pace did not overwhelm me. & 3.06 & 0.90 \\
I would play the game again. & 3.35 & 1.32 \\
\bottomrule
\end{tabular}
\caption{Results of the items in the category Satisfaction.\\
Mean values and standard deviations of participant ratings.}
\label{t3}
\end{table}

The participants had to answer 6 items in the area of satisfaction. With a maximum possible score of 85 points per item, the total score was 510. 327 out of 510 points were achieved, which corresponds to a satisfaction rate of 64\%. The results of all three components (effectiveness + efficiency + satisfaction) are used to measure the general usability of the PCLQ digital learning game. Usability is therefore 63.33\%. In the following chapter, the results of the two data surveys are placed in the context of the EDUMING concept. 

\section{DISCUSSION}
With a total score of 73 (M = 4.29), the main mechanic of the game, namely which chip is currently needed to solve the task, was rated the most highly. With a small variance (s = 0.77), there was agreement among the participants on this point. Other mechanics of the game, however, did not seem immediately clear. For example, the question "I could immediately understand how the game's mechanics work." received only a few points (P = 38; M = 2.24). The same applies to the question, "I knew what I had to do in the game." (P = 41; M = 2.41). The standard deviation for both questions was high (>1.3), indicating that the participants' opinions differed greatly and there was no uniform understanding. This suggests that the participants had different experiences. The mechanics of picking up letters (P = 49; M = 2.88) and discarding them (P = 44; M = 2.53) also received few points and were therefore viewed rather critically. Although an option in the main menu explaining the game's principle was available, it was apparently not activated. The problem here lies in a design decision by the developer. The explanation of the mechanics is tied to a representation of the assistant. Even when the assistant moves back and forth, there is no text, so it was not perceived as a selectable option. In the game itself, picked up letters could also be discarded if they were not relevant to the current task. To improve this mechanic, another station, randomly generated in the arena, is to be added where picked up voice chips can be exchanged for others. This adjustment should improve the flow of the game. Navigation in the game also falls fairly close to neutral (P = 46; M = 2.71) and exhibits a high standard deviation (s = 1.45). The high variance in some individual items may be explained by the results of the first survey. Most participants had little experience with digital games, either in their studies or in their personal lives. Some participants, however, had no problems navigating the game. The data confirm this assumption.

The movement of the actual game piece was rated better (P = 61; M = 3.59). However, the spread (s = 1.28) also shows different experiences here. As the educational game was mainly tested on smartphones and was controlled using two joysticks, it was more difficult for less experienced participants to navigate through the arena. The data shows that user guidance needs to be improved so that basic mechanics are better understood. Even if explanations are provided, they need to be more visible. The participants agreed that the tasks were perceived as easy (P = 61; M = 3.59). This was to be expected, as the game and the learning mechanics implemented are intended to promote spelling skills. The fact that the rating is nevertheless rather low is probably due to the underlying mechanics. For example, the question “It was easy to complete the tasks with minimal effort.” was rated low (P = 46; M = 2.71). The tasks were simple, but difficult to solve due to the mechanics. This can probably be explained by the navigation in the game via smartphone, where letters that are not needed do not have to be removed by right-clicking with the mouse, but via an icon. 

This is explained in the general game instructions. As already mentioned, many participants could not find them. PCLQ is an arena shooter. The game style and the implemented mechanics probably led to overtaxing the less experienced participants (P = 45; M = 2.65). In contrast, the game pace was perceived as less demanding (P = 52; M = 3.06). The previously mentioned emphasis on the existing explanation of the game's progression is intended to counteract overtaxing in the future. The visuals used in the game (P = 68; M = 4.06) and the music (P = 69; M = 4.06) captured the participants' attention and were rated very highly. The size of the playing field was also good and not overloaded with too many GUI elements (P = 69; M = 4.06). However, the variance here was somewhat larger (s = 1.14). The overall design and the musical accompaniment of the game impressed the participants. The question "I had no problems starting the game." was rated moderately to slightly positively (P = 51; M = 3.00). This can be explained by how the participants had to start the application. Because the app was still in development, it was not available via a store (e.g., Google Play), but was hosted by GX Games, which meant that the game had to be loaded first. Since the internet connection at the university is slow, the game's small file size (45 MB) led to longer waiting times before the application could even be launched. The comprehensibility of the overall concept was rated slightly positively (P = 56; M = 3.24). This value could be improved by emphasizing the explanation of the general mechanics. The timely recording of the voice chips dropped by the robot vacuums was rated just above "neutral" (P = 52; M = 3.06) with a relatively high standard deviation (s = 1.39). This can be adjusted by increasing the corresponding variable, specifically the time interval.

The individual adaptation of such mechanics in the game, depending on the learning group in which it is used, is an important aspect of the EDUMING concept, entirely in the spirit of constructionism, and a decisive advantage over other approaches in this field. Displaying the currently required letter, i.e., the task currently to be solved, helped participants react accordingly (P = 57; M = 3.35). The basic orientation in the game is good. Regardless of this, some participants had very different starting and navigation experiences. As already mentioned, this can be explained by the different experiences with digital games, the poor internet connection, and the less than optimal instructions regarding the game mechanics. The questions "The game was fun" (P = 57; M = 1.32) and "I would play the game again" (P = 57; M = 1.32) were answered almost identically. Both questions showed a high degree of variance (s = 1.32), which suggests that some found the game entertaining and others not. The game character, Professor Chips, was rated neutrally (P = 52; M = 3.00), meaning that he is perceived as neither positive nor negative. To improve this, additional animations will be integrated so that the character reacts more to the game world. Randomly generated comments from the character are also possible to motivate the player. The question "The game was easy to play" (P = 48 points; M = 2.82) was rated below the neutral mark (s = 1.13). Although the actual tasks in the game were simple, this can be explained by the game mechanics, which were not so quickly understood. Overall, the participants considered the game design to be successful. There is room for improvement in the areas of core mechanics and handling to avoid confusing or overwhelming users. The overall fun factor should be improved through additional mechanics in the game.

The usability of the digital learning game PCLQ was rated at 63\%, which is a good result despite the very different experiences of the participants. However, the adaptations already mentioned are intended to improve this value. The results show that digital learning games developed according to the EDUMING concept have the potential to be used in the classroom. The results of the first survey showed that prospective teachers have a positive attitude towards these media and, if the conditions are right, are willing to use digital learning games in their lessons. The EDUMING concept is intended to simplify the framework for implementation. Specifically in terms of finance through the use of free tools (further development with GameMaker Studio 2 and publication via GX Games), but also in terms of the time and organizational effort. This means that the games can also be adapted together with the learners to convey the current lesson topic.

The study provides concrete recommendations for teachers, researchers, and developers. Thanks to the easy adaptability of game templates, EDUMING offers teachers a flexible way to implement game-based learning even in heterogeneous learning groups. This approach is particularly suitable for project-oriented learning units, as learners can not only learn through play but also be creative on their own. Developers can create additional templates to cover common curricular topics. The more templates available, the easier it is to use EDUMING in learning situations. In the interests of further research, future studies should examine long-term competency development and changes in motivation among learners. Long-term studies that track EDUMING games over an entire school year could provide meaningful results and help to further empirically substantiate the pedagogical benefits of constructionism.

\section{CONCLUSION}
Video games have become an increasingly important leisure activity and play a crucial role in the lives of young people \cite{15}. It is known that video games can increase motivation in learning \cite{16}. Nevertheless, these media are used in this context to a limited extent. The results of the initial data collection show that future teachers, despite having little experience in this area, are open to the use of digital learning games in the classroom and see their added value compared to analog learning methods. Barriers to the use of this technology include the varying framework conditions (e.g., technical equipment, stable internet connection) at schools.

If teachers do not receive support from school administration, it reduces the likelihood that digital learning games will be used in the classroom. As a generational shift takes place in schools in the coming years, attitudes towards digital learning games as an important medium for developing skills will also change. To improve the framework for its use, the EDUMING concept was developed as a new approach to close a gap previously ignored by established concepts (gamification, game-based learning, and serious games). This gap builds on the theory of constructionism developed by Papert. A game developed according to the EDUMING concept can not only be used directly but can also be further developed by users or learners to potentially create something completely new.

The optimized games can then be shared with others, played, and discussed regarding possible improvements. The entire process, from the mere use of an EDUMING game to its adaptation, serves the purpose of skills development and is intended to promote the transition from purely passive use to active self-development. Testing of the PCLQ game, developed according to the EDUMING concept, has already shown good results and was rated with a usability of 63\% despite the participants' varying gaming experiences. To further improve usability, the mechanics will be further optimized. After adaptation, the digital learning game will be made available via a GitHub repository with neutral visualizations, i.e., without the PCLQ design. There are plans to adapt, test, and make available other GameMaker Studio 2 templates similarly. The goal of confirming the acceptance and usability of the PCLQ game, developed according to the EDUMING concept, has been empirically proven. In a next step, the concept will be used to provide empirical evidence for learners' learning progress and/or competency development. The task mechanics currently integrated into PCLQ form the basis for investigating formative learning processes.

\section{ACKNOWLEDGMENTS}
The author(s) received no financial support for the research, authorship, and/or publication of this article.

\clearpage
\renewcommand{\refname}{REFERENCES} 
\bibliographystyle{unsrt}
\bibliography{references} 

\end{document}